# The triennial International Pigment Cell Conference (IPCC)

Neil F. Box [%], Lionel Larue, Prashiela Manga, Lluis Montoliu, Richard A. Spritz, Fabian V. Filipp [*]

**Correspondence**

[*] Fabian V. Filipp, University of California Merced, Systems Biology and Cancer Metabolism, Program for Quantitative Systems Biology, 5200 North Lake Road, Merced, United States, CA 95343;

[%] Neil Box, University of Colorado Denver, Department of Dermatology and Epidemiology, United States, CO, Aurora;

Telephone: 858-349-0349 and 303-724-0160; [*] [%] Email: filipp@ucmerced.edu and neil.box@ucdenver.edu

**Summary**

The International Federation of Pigment Cell Societies (IFPCS) held its XXIII triennial International Pigment Cell Conference (IPCC) in Denver, Colorado in August 2017. The goal of the summit was to provide a venue promoting a vibrant interchange among leading basic and clinical researchers working on leading-edge aspects of melanocyte biology and disease. The philosophy of the meeting, entitled Breakthroughs in Pigment Cell and Melanoma Research, was to deliver a comprehensive program in an inclusive environment fostering scientific exchange and building new academic bridges.

**Significance**

This document provides an outlook on the history, accomplishments, and sustainability of the pigment cell and melanoma research community. Shared progress in the understanding of cellular homeostasis of pigment cells but also clinical successes and hurdles in the treatment of melanoma and dermatological disorders continue to drive future research activities. A sustainable direction of the societies creates an international forum identifying key areas of imminent needs in laboratory research and clinical care and ensures the future of this vibrant, diverse and unique research community at the same time.

Important advances showcase wealth and breadth of the field in melanocyte and melanoma research and include emerging frontiers in melanoma immunotherapy, medical and surgical oncology, dermatology, vitiligo, albinism, genomics and systems biology, precision bench-to-bedside approaches, epidemiology, pigment biophysics and chemistry, and evolution. This report recapitulates highlights of the federate meeting agenda designed to advance clinical and basic research frontiers from melanoma and dermatological sciences followed by a historical perspective of the associated societies and conferences.

**Introduction—Program highlights of the XXIII IPCC**

The XXIII IPCC encompassed a series of opening and award lectures, including the IFPCS Presidential lecture and honorary lectures by the recipients of the Fitzpatrick, Lerner, and Seiji Awards. These awards are bestowed by regional pigment cell societies to leading international scientists who have made extraordinary contributions to pigment cell research. There were 8 plenary sessions and 32 concurrent sessions accomplishing a balance between clinical and basic melanoma and dermatology research.

In addition to opening lectures, the first day included a plenary session on the evolution of pigmentary systems across the animal kingdom and three concurrent sessions on advances in vitiligo, melanocyte, and melanoma biology as well as disease management. On the second day, two plenary sessions covered fascinating advances in the genetics of coloration in various animal species and in UV light responses and DNA repair. The fourth day included two plenary sessions and four concurrent sessions that emphasized basic biology of melanoma and the latest developments in translational melanoma therapies and resistance. Day five was headlined by a plenary session with keynote speakers on the impact of regenerative medicine on pigmentary diseases and a plenary session on the immunologic control of melanocytes and melanoma. General sessions covered evolutionary biology of pigmentation, developmental biology, genetics, genomic profiling, neuroscience and pigmentation, melanocyte stem cells and regenerative medicine, hair biology, mouse and non-mouse models of pigmentation, melanosome biogenesis and transfer, melanin function and chemistry, and UV signaling in the melanocyte. The program also included three lunchtime poster sessions with presentation of about 200 posters. Importantly, sessions were specifically dedicated to provide mentoring and career opportunities for young scientists, women in science, and underrepresented minorities in science.

The basic and clinical melanoma research sessions presented a state-of-the-art picture of immunotherapy, precision medicine, and genomic profiling of skin cutaneous, uveal, mucosal, and acral melanoma [1]. Keynote speaker Dr. Jeffrey Weber highlighted progress in melanoma management, while keynote speaker Dr. Douglas Brash presented the latest developments in DNA damage and melanoma risk with his work on UV-induced DNA damage and chemiexcitation of melanin. Frontiers of basic melanoma biology were introduced by keynote speaker Dr. Richard Marais who presented advances in targeted melanoma research integrating mouse models and therapies. The basic and clinical dermatology research sessions covered pigmentary processes, diseases, and treatments. The program featured keynote speakers including Dr. Sarah Tishkoff who addressed the genetic basis for skin pigmentation in African populations [2, 3]. Dr. Hopi Hoekstra identified the molecular basis of parental care evolution and mammalian striping phenotypes uncovered using the mouse as a model [4]. Dr. Rudolf Jaenisch provided insights into the forefront of mammalian embryonic and induced pluripotent stem cell research demonstrating the fundamental rules of cellular reprogramming [5]. Dr. Dennis Roop discussed progress on differentiating human induced pluripotent stem cells into keratinocytes, fibroblasts, and applications for skin grafting including melanocytes [6].

Highlights included updates on precision dermatology, bench-to-bedside approaches, genome-wide association studies, genetic diversity of pigmentation, and progress in immunotherapy [7]. Dr. Ian Jackson presented the analysis of the UK Biobank cohort of over 0.5 million genotyped individuals including detailed phenotype self-reports on pigmentation and hair color. This analysis allowed construction of detailed polygenetic models for blonde and red hair color, and penetrance assessment of many of the low frequency melanocortin 1 receptor (*MC1R*) variants. Dr. Nicolas Hayward delivered analysis of whole-genome sequences from cutaneous, acral and mucosal subtypes of melanoma. In contrast to frequently occurring coding and non-coding mutations in cutaneous melanoma, the Australian consortium resolved structural changes and novel signatures of mutagenesis in acral and mucosal melanomas, not previously identified in melanoma [8].

A number of presentations also explored the genetic diversity of pigmentation in non-human and non-mouse systems. Dr. Cheng-Ming Chuong presented a novel foray into the world of feather pigment pattern formation, where color patterns important for animal behavior and speciation is modulated by the presence, arrangement, or differentiation of melanocytes [9]. Among many other important clinical advances that were introduced during the meeting, Dr. Robert Andtbacka presented final results of a phase II multicenter combination trial that included ipilimumab and oncolytic virus immunotherapy [10]. Teams of Dr. Frank Meyskens and Dr. Stephane Rocchi elucidated the pivotal role of metabolic and transcriptional reprogramming in the switch of melanoma cells toward an invasive and drug-resistant phenotype [11, 12]. Dr. Jeffrey Weber and Dr. Dirk Schadendorf pointed to the importance of cancer-germline antigens that predict resistance to therapy response [13].

*Historical perspective of IFPCS and PASPCR unfolding the purpose of the IPCC—mission, vision, and future needs*

The International Pigment Cell Conference has emerged as the sole international scientific convention devoted to the study of the normal pigment cell and the advancement of basic, translational, and clinical research on diseases involving pigment cells. The first meeting was held in New York in 1946 and reconvenes on a triennial basis (Figure 1). Recent sites for this important international gathering have been Singapore (2014), Bordeaux (2011), and Sapporo (2008) (Table 1). While each meeting has had its own unique focus, commonalities for over 70 years have always been melanin, pigment cells and melanoma.

In 1977, the International Federation of Pigment Cell Societies (IFPCS) was developed first as the International Pigment Cell Society and then formalized as IFPCS in 1990. With the IFPCS formally acting as a means to provide interaction between the independently formed sister societies, the IPCC meeting was chartered as a main function of IFPCS. The Japanese Society for Pigment Cell Research (JSPCR) was founded in 1984, the European Society for Pigment Cell Research (ESPCR) in 1985, the Pan-American Society for Pigment Cell Research (PASPCR) in 1988, and the Asian Society for Pigment Cell Research (ASPCR) in 2004, each becoming members of IFPCS at that time. Thus, the mission of the IFPCS is to disseminate cutting edge research via the publication of the federation's journal, *Pigment Cell and Melanoma Research*, and to host the IPCC meeting. The IPCC remains the primary vehicle to promote worldwide scientific interchange for those international investigators who use an array of approaches to study pigment cell function in normal biology and disease. An international forum to present data, to discuss ideas, to identify key areas of needed research and to set new research directions is essential to ensuring the future of this vibrant, diverse and unique research community. The success of the IPCC to play this role in the pigment cell community is evidenced by the fact that attendance has been on the rise (Figure 2).

*Recapitulation of the XXIII IPCC*

The XXIII IPCC, representing the four member Pigment Cell Societies of the International Federation, was hosted by the Pan-American Society for Pigment Cell Research in Denver, Colorado, on August 26-30, 2017. 540 registrants from 28 countries participated in the 8 plenary and 32 concurrent sessions. 8 keynote speakers led the program, along with 21 plenary session speakers and 80 invited speakers. A further 103 speakers were chosen from submitted abstracts. The main program was complemented by four pre-conference satellite symposia and four pharmaceutical company sponsored symposia. The annual meeting of the Melanoma Prevention Working Group (MPWG) was hosted for the first time at the IPCC meeting. The MPWG functions under the umbrella of the Southwest Oncology Group (SWOG) and the Eastern Cooperative Oncology Group (ECOG). The MPWG meeting promoted clinically relevant research focused on key aspects of melanoma prevention and discussed innovative paths forward into clinical trials of chemoprevention agents for high-risk melanoma patients [14]. The comprehensive scientific program of the IPCC 2017 contained an equal weight of basic and clinical research in pigmentary diseases and melanoma. The importance of studying pigment cells is defined most dramatically by recent advances in melanoma therapies that can be lifesaving, but also by the many diseases and conditions intersecting with the pigmentary system that remain in need of effective treatments.

A major program highlight was a dedicated session on advances in basic and clinical research on albinism. Albinism is a rare genetically inherited condition affecting approximately one in twenty thousand people in most world populations. Albinism occurs at relatively high frequency in sub-Saharan Africa. People with albinism experience visual deficits, hypopigmentation, and sun-sensitivity. In less advanced societies, people with albinism may not have access to healthcare specialists, sunscreen, or protective clothes. Moreover, in sub-Saharan Africa, a humanitarian crisis is presently underway affecting patients with this disease [15]. The dedicated session on albinism raised awareness to this important issue, and sent a message of solidarity to patients suffering from albinism, leaving a unique footprint in the field of pigment cell biology.

*Planning and Organization of IPCC 2017—Special format, challenges, sustainability*

IPCC 2017 was planned for late summer, August 26-30, 2017 in downtown Denver, Colorado. The strategy for designing and marketing of IPCC 2017 was informed by multiple lines of reasoning. First, with new melanoma therapies providing improvements in medical care, the field has progressed in an increasingly clinical direction, in some ways diminishing the focus on the basic research that has given rise to successful treatment strategies. Inevitably, new strategies will be needed to further advance the treatment of melanoma patients, and without a strong basic research base, these advances will not be possible. Moreover, successes in the translational arena of melanoma need to be duplicated in other key areas of pigment cell research, including vitiligo, melasma, and other pigmentary diseases. Providing a strong conference venue to support basic research in pigmentary diseases and melanoma in essence embraced the true mission of the IPCC meeting and of the IFPCS organization. A second influencing factor in designing IPCC 2017 was the recognition that with the World Melanoma Congress (WMC) associated with the Society for Melanoma Research (SMR) being scheduled for Brisbane in October, the IPCC would be the only major melanoma meeting to be held in the Northern Hemisphere in 2017. Thus a strong focus on melanoma would allow North American clinicians and researchers proximal access to the latest advances in the field. The third factor embraces all who are interested in pigment cell biology and disease by providing a comprehensive format for promoting interchange between such diverse groups of researchers. Thus, the theme of IPCC 2017 was strategically chosen as *Breakthroughs in Pigment Cell and Melanoma Research*. In essence, IPCC 2017 was at once an outreach to all members of the community and a call to stand unified in facing potential funding challenges as a field.

Initially, a dual focus on pigmentary disease and melanoma seemed to allow maximal opportunity for industry based meeting support. With strong support from the melanoma committee, major pharmaceutical company support was obtained to support the melanoma program. However, it proved more challenging to raise support from companies with an interest in pigment cell function and disease. This was compensated for by outstanding institutional support from groups within the University of Colorado system affiliated with the local organizing committee. Consequently, the meeting was supported by the melanoma related pharmaceutical industry, the

University of Colorado, by conference registration fees and even through generous support from speakers, session chairs and attendees.

## Conclusions and Outlook

*IFPCS and the IPCC as its global forum*

The IFPCS is a vital global umbrella organization for the regional pigment cell societies. Over more than seven decades, the triennial IPCC has been an opportunity for scientist across the globe to gather to discuss recent findings, progress, and ongoing research. Consistently, the federation and local societies have provided the international research community with essential resources on all aspects of pigment cells including development, cell and molecular biology, genetics, diseases of pigment cells including melanoma.

**Declarations**

**Competing interests**

There is no competing financial interest.

**Acknowledgements**

We are grateful to support by the councils of the International Federation of Pigment Cell Societies (IFPCS) and the Pan-American Society for Pigment Cell Research (PASPCR). Conference information along with all meeting abstracts of the XXIII triennial International Pigment Cell Conference (IPCC) of the International Federation of Pigment Cell Societies (IFPCS) and the XIV annual Society of Melanoma Research (SMR) is is available at https://doi.org/10.1186/s12967-018-1609-1, https://doi.org/10.1111/pcmr.12622, and https://doi.org/10.1111/pcmr.12656.

**Funding**

The councils of the International Federation of Pigment Cell Societies (IFPCS) and the Pan-American Society for Pigment Cell Research (PASPCR) are grateful for support by grant R13 AR071775 from the National Institutes of Health.

**Availability of preprint publication**

The manuscript was made publically available to the scientific community on the preprint server arXiv at https://arxiv.org/abs/1808.05755.